\documentclass[twoside,twocolumn,9pt]{article}
\usepackage{extsizes}
\usepackage[super,sort&compress,comma]{natbib} 
\usepackage[version=3]{mhchem}
\usepackage[left=1.5cm, right=1.5cm, top=1.785cm, bottom=2.0cm]{geometry}
\usepackage{balance}
\usepackage{widetext}
\usepackage{times,mathptmx}
\usepackage{sectsty}
\usepackage{graphicx} 
\usepackage{lastpage}
\usepackage[format=plain,justification=justified,singlelinecheck=false,font={stretch=1.125,small,sf},labelfont=bf,labelsep=space]{caption}
\usepackage{float}
\usepackage{fancyhdr}
\usepackage{fnpos}
\usepackage[english]{babel}
\usepackage{array}
\usepackage{droidsans}
\usepackage{charter}
\usepackage[T1]{fontenc}
\usepackage[usenames,dvipsnames]{xcolor}
\usepackage{setspace}
\usepackage[compact]{titlesec}
\usepackage{url}
\usepackage{xcolor}

\usepackage{epstopdf}

\definecolor{cream}{RGB}{222,217,201}

\begin{document}

\pagestyle{fancy}
\thispagestyle{plain}
\fancypagestyle{plain}{

}

\makeFNbottom
\makeatletter
\renewcommand\LARGE{\@setfontsize\LARGE{15pt}{17}}
\renewcommand\Large{\@setfontsize\Large{12pt}{14}}
\renewcommand\large{\@setfontsize\large{10pt}{12}}
\renewcommand\footnotesize{\@setfontsize\footnotesize{7pt}{10}}
\makeatother

\renewcommand{\thefootnote}{\fnsymbol{footnote}}
\renewcommand\footnoterule{\vspace*{1pt}%
\color{cream}\hrule width 3.5in height 0.4pt \color{black}\vspace*{5pt}} 
\setcounter{secnumdepth}{5}

\makeatletter 
\renewcommand\@biblabel[1]{#1}            
\renewcommand\@makefntext[1]%
{\noindent\makebox[0pt][r]{\@thefnmark\,}#1}
\makeatother 
\renewcommand{\figurename}{\small{Fig.}~}
\sectionfont{\sffamily\Large}
\subsectionfont{\normalsize}
\subsubsectionfont{\bf}
\setstretch{1.125} 
\setlength{\skip\footins}{0.8cm}
\setlength{\footnotesep}{0.25cm}
\setlength{\jot}{10pt}
\titlespacing*{\section}{0pt}{4pt}{4pt}
\titlespacing*{\subsection}{0pt}{15pt}{1pt}

\fancyfoot{}
\fancyhead{}
\renewcommand{\headrulewidth}{0pt} 
\renewcommand{\footrulewidth}{0pt}
\setlength{\arrayrulewidth}{1pt}
\setlength{\columnsep}{6.5mm}
\setlength\bibsep{1pt}

\makeatletter 
\newlength{\figrulesep} 
\setlength{\figrulesep}{0.5\textfloatsep} 

\newcommand{\topfigrule}{\vspace*{-1pt}%
\noindent{\color{cream}\rule[-\figrulesep]{\columnwidth}{1.5pt}} }

\newcommand{\botfigrule}{\vspace*{-2pt}%
\noindent{\color{cream}\rule[\figrulesep]{\columnwidth}{1.5pt}} }

\newcommand{\dblfigrule}{\vspace*{-1pt}%
\noindent{\color{cream}\rule[-\figrulesep]{\textwidth}{1.5pt}} }

\makeatother

\twocolumn[
  \begin{@twocolumnfalse}
\vspace{3cm}
\sffamily
\begin{tabular}{m{9cm} p{9cm} }

\LARGE{\textbf{Effect of geometry on the dewetting of granular chains by evaporation}} \\
\vspace{0.3cm} & \vspace{0.3cm} \\

\large{Cesare M. Cejas$^{\ast}$\textit{$^{a}$}$^{\dag}$ Lawrence A. Hough,\textit{$^{a}$} Christian Fr\'etigny,\textit{$^{b}$} and R\'emi Dreyfus$^{\ast}$\textit{$^{a}$}} \\

\normalsize{Understanding evaporation or drying in granular media still remains complex despite recent advancements. Evaporation depends on liquid transport across a connected film network from the bulk to the surface. In this study, we investigate the stability of film networks as a function of the geometry of granular chains of spherical grains. Using a controlled experimental approach, we vary the grain arrangement or packing and measure the height of the liquid film network during evaporation as packing shifts from loose-packed to close-packed arrangement. This height can be calculated from an equilibrium between hydrostatic pressure and the capillary pressure difference in the vertical film network. Following a simulation approach using Surface Evolver, we evaluate the pressure variation due to dewetting of the meniscus volume in the grains in both the percolating front and evaporating front within the two-phase zone of air/water mixture. Results show good agreement between model and experiment. We find that above a ``critical'' packing angle, the liquid continuity is broken and films connections fragment into separate, isolated capillary bridges.} \\

\end{tabular}

 \end{@twocolumnfalse} \vspace{0.6cm}

  ]

\renewcommand*\rmdefault{bch}\normalfont\upshape
\rmfamily
\section*{}
\vspace{-1cm}

\footnotetext{\textit{$^{a}$~Complex Assemblies of Soft Matter, CNRS-Solvay-UPenn UMI 3254, Bristol, PA 19007-3624, USA}}
\footnotetext{\textit{$^{b}$~Sciences et Ing\'enierie de la Mati$\grave{e}$re Molle CNRS SIMM UMR 7615 ESPCI, 10 rue Vauquelin, Paris 75005 France}}
\footnotetext{\dag~Microfluidics, MEMS, Nanostructures Laboratory, CNRS Gulliver UMR 7083, Institut Pierre Gilles de Gennes, PSL Research University, 6 rue Jean Calvin, Paris 75005 France}





\section{Introduction}

Water evaporation out of porous media affects mechanical and biological properties of soil and  understanding its mechanism is significant in hydrology~\cite{Lehmann08}. Previous studies~\cite{Lehmann08, Shokri08, Yiotis12b, Prat99} have tackled the challenge of predicting the evaporation rates and drying phenomenon. Some studies~\cite{Dreyfus14} have used coupled evapotranspiration (combined effects of evaporation from porous media and transpiration from leaves) rates to predict the total amount of water lost in porous media in the presence of root systems.  It is now well-established that the evaporation rate is intricately related to a so-called Partially Saturated Zone (PSZ)~\cite{Cejas17} or vadose zone, which is an air-liquid mixture that forms beneath the surface and made of connected liquid films and disconnected droplets~\cite{Cejas17, Shokri09}. However, what affects the organization and geometrical characteristics of this layer is still under debate~\cite{Prat02, Prat07}.

The study of liquid flow through pores in the PSZ has been described by several approaches. In one approach, the liquid transport through the PSZ is explained by the pore polydispersity. Pore size distribution is the key point that induces a pressure difference that permits liquid flow from larger pores into smaller ones ~\cite{Lehmann08, Shokri08}. In another approach, individual pore shape is what controls the transport of liquid~\cite{Dong95, Chauvet09, Prat07, Chauvet10}, particularly the presence of sharp corner edges that influence evaporation rates. Experiments and simulations both show that corners within the pore walls help maintain film stability ~\cite{Chauvet09, Yiotis12b, Prat07, Dong95}. Films that carpet these corner edges or gutters~\cite{Fuerstman07} have a pressure gradient along their length and thus act as hydraulic pathways for easy and rapid transport of water from the bulk (interior of the porous medium) to the surface. Recent results~\cite{Chen18} have demonstrated hydraulic connectivity from chains of liquid bridges proving that small quantities of fluid can have a huge impact on the transport properties in granular materials~\cite{Herminghaus05, Reiser15}. All these research highlight the fact that drying in porous media is complex. As different approaches subsist, certain aspects have yet to find sufficient explanations.

It has also been demonstrated that evaporation out of a complex 2D/3D porous medium can be analogous to the evaporation mechanism out of a polygonal capillary tube (e.g. square), opening new routes towards a simplified and accurate description of evaporation ~\cite{Prat07, Yiotis12b, Yiotis12a}. In both cases, evaporation mechanisms involve two regimes: (1) a rapid first regime due to hydraulic connections pinned to the corner that connect the bulk to the surface, and; (2)  a slow second regime due to depinning of these films from the corner and the appearance of a dry zone through which water diffuses in the gas phase~\cite{Camassel05, Chauvet09, Yiotis12a, Yiotis12b, Cejas17}. Similarities also exist in terms of physical orgin. As evaporation occurs in porous media such as soil, the menisci between particles closer to the soil surface have higher curvatures that create a pressure gradient responsible for water flow. During evaporation in square tubes, the films pinned to the corner also become thinner and the pressure difference along the film length drives water flow. The sharp geometrical features at the contact between particles in soil is analogous to the sharp corners of the square tube. These features help disrupt or maintain liquid film stability~\cite{Yiotis12a, Yiotis12b} and affect drying kinetics. The morphology of the corner films changes with pore size and shape~\cite{Prat07} and pore shape is influenced by granular arrangement. To further improve our understanding of the evaporation process, it is necessary to examine the stability of liquid films in relation to the geometry of a very simple model made of a granular chain.

In this paper, the main objective is to demonstrate the influence of grain arrangement on liquid film stability and formation, as the liquid gradually undergoes dewetting by evaporation. We show that evaporation dynamics are controlled by the stability of the hydraulic connections and that packing configurations of the grains contribute to the transition from capillary-supported (fast) to diffusion-controlled (slow) regime. Experimentally, we control the conformation of the beads in the tube to vary configuration from `'straight packing'' to `'close-packing''. We then use simulations to describe the shape and form of the liquid film attached to the grains and then compare these results with experiments.

\section{Experimental Set-up}

The experimental set-up consists of two types of borosilicate capillary tubes (Vitrocom, NJ, USA) and one type of glass spheres of diameter $d = (1.0\pm 0.2$)~mm (Sigma Aldrich, USA). The spheres are carefully sieved to closely select monodisperse sizes. The tubes have lengths of $L = 10$~cm and circular cross-sections. The tubes are filled with glass spheres to resemble a ``single-column" porous medium. Given monodisperse glass spheres, two different internal diameters (1.5~mm, 2.0~mm) of the circular capillary tube are used to ensure specific packing configurations. The beads inside the circular tube form a 2D monolayer packing, more appropriately described as a ``granular chain'' since it lacks the characteristics of true granular packing (e.g. complex pore networks, disorder). The confinement of spheres in a cylindrical tube, especially when the former is smaller than the latter, may result to a break in chiral symmetry and formation of spontaneous twisting conformations~\cite{Pickett00, Fu16}, e.g. helices. However, we observe in all our experiments that the bottom of the PSZ only reaches a maximum depth of 2.6~cm from the surface. Within this depth, we observe no helical structures, the packing stays achiral, and the zigzag arrangement of the spheres is planar. If helical structures are present, they do not concern the PSZ and are found within the fully wet zone for the duration of the experiment. The inner tube walls are made hydrophobic using a silane solution (OMS Solutions, Canada) to reduce wetting effects on the capillary wall.

The columns are vertically mounted on a stand and the top end remains open. They are fully saturated with deionized water using a vacuum pump to eliminate the presence of bubbles. The glass spheres, pre-washed with 0.1~M HCl, have a constant wetting angle $\theta = (17 \pm 5)^{\circ}$. The spheres are placed in the capillary tubes in different packing configurations defined by the packing angle $\alpha$, as shown in Fig.1a-c. Straight configurations have a packing angle of $\alpha = 180^{\circ}$, while loose configurations have angles of $60^{\circ} < \alpha \leq 180^{\circ}$, and close-packed configurations have angles of $\alpha \sim 60^{\circ}$. The value of $\alpha$ is the average angle measurement of the beads.

The experimental porosities are measured using the imbibition method~\cite{Anovitz15}, where we take the masses of the empty system with respect to the same system filled with water. The porosities can also be theoretically calculated using a geometrical argument and we found no significant difference between both methods. This agreement strongly suggests that the void spaces inside the tube are properly filled with water.

All experiments are carried out in a controlled environment chamber (Electro-Tech Systems Model $5100$, PA, USA). The chamber has a fan that periodically turns on and off. The relative humidity and temperature inside the chamber are kept constant at $(20.0 \pm 2.0)\%$ and $(32\pm 2)^{\circ}$C respectively.  We take images of the receding front with time using Canon $500$D SLR camera with $18$ to $55$~mm lens with a resolution of 35 $\mu$m per pixel. Experiments for each configuration are reproduced for a minimum of 3 times.

\begin{figure}[h]
\centering
\includegraphics[width=3in]{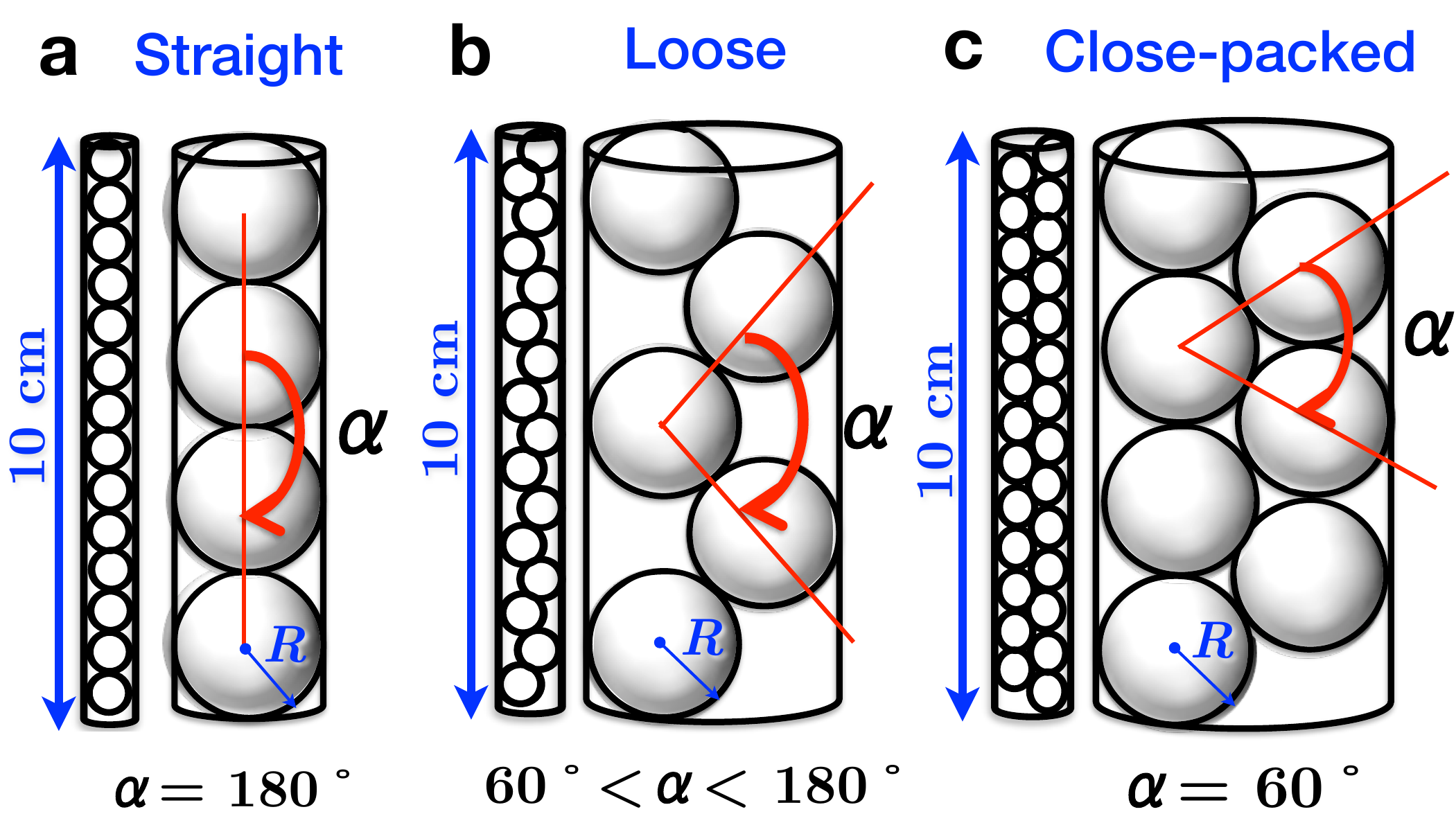}
\caption{Set-up of ``single-column" porous media or ``granular chain'' configurations of circular tubes filled with glass spheres of packing angle, $\alpha$. (\textbf{a}) Straight configuration with $\alpha = 180^{\circ}$. (\textbf{b}) Loose-packed configuration with $60^{\circ} < \alpha < 180^{\circ}$. (\textbf{c}) Close-packed configuration with $\alpha \sim 60^{\circ}$.}
 \label{Fig1}
\end{figure}


\section{Experimental Results}

\begin{figure}[h]
\centering
\includegraphics[width=3in]{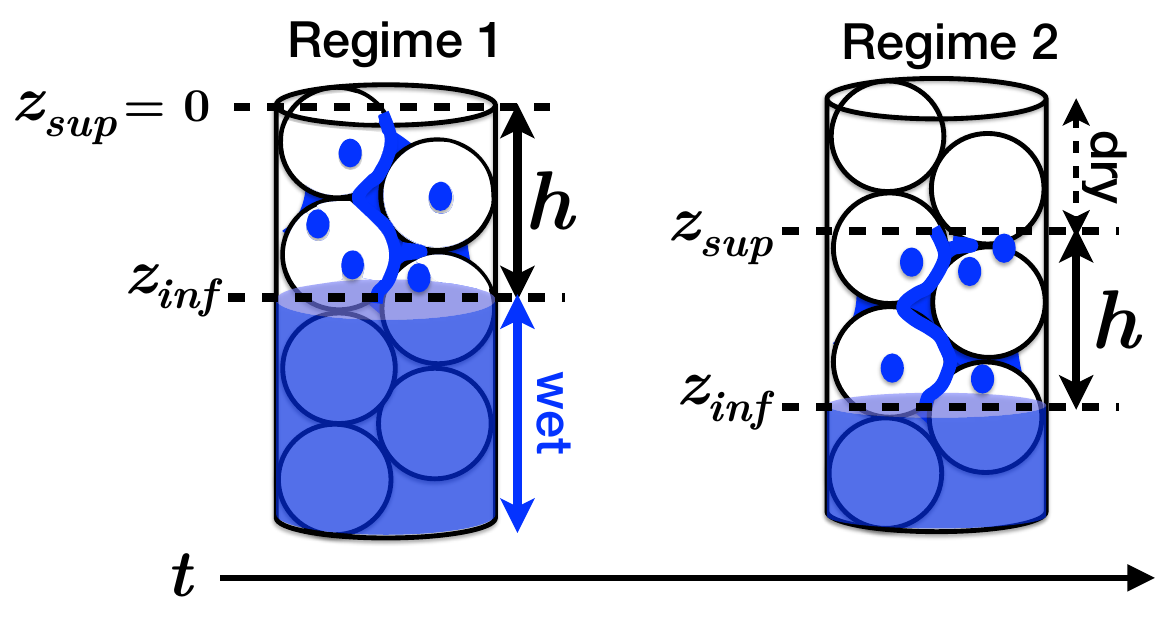}
\caption{Schematic of the two evaporation regimes, showing $z_{inf}$ (percolating front), $z_{sup}$ (evaporating front), and $h$ is the characteristic length or height of the PSZ. In regime 1, we observe the fully wet zone and PSZ; while in regime 2, we observe the fully wet, PSZ, and dry zones.}
\label{Fig2}
\end{figure}

\begin{figure*}[h]
\centering
\includegraphics[width=6.5in]{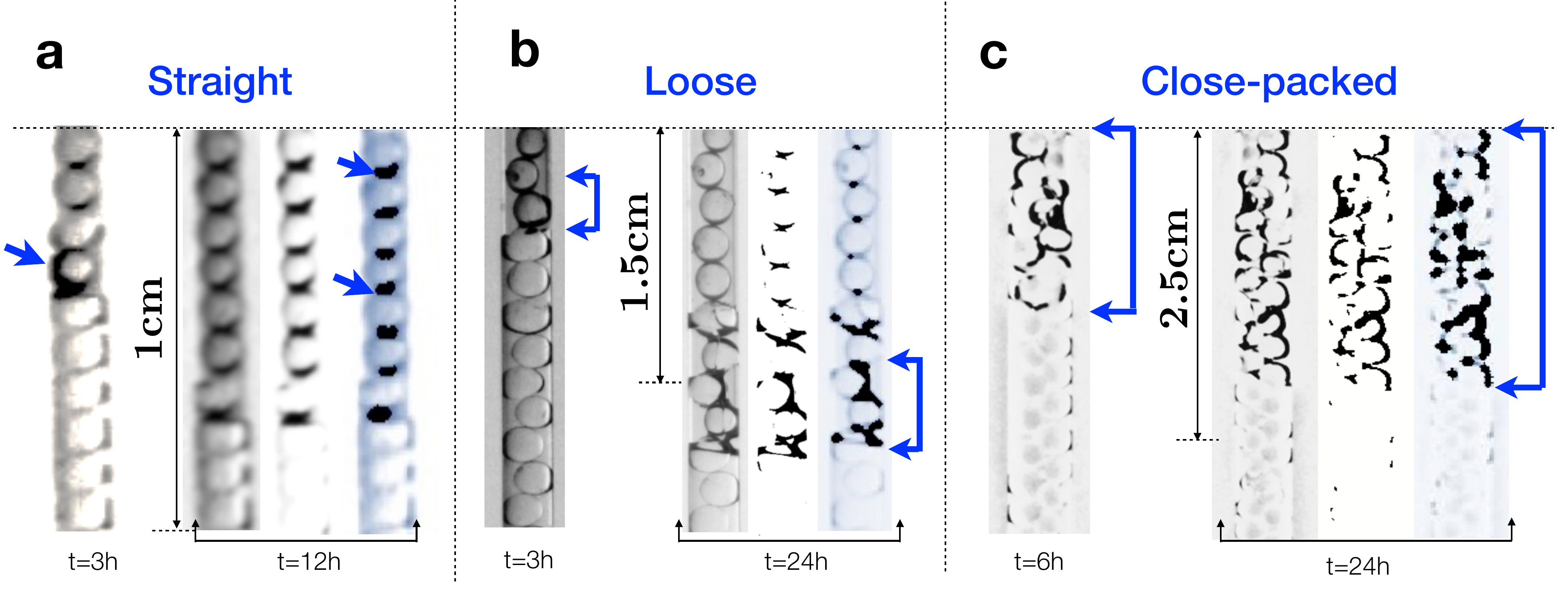}
\caption{Experimental images of PSZ for (\textbf{a}) straight, (\textbf{b})  loose, and (\textbf{c}) close-packed systems. The images are treated using Matlab to distinguish the films from the beads. The arrows indicate liquid films emphasized by darker regions. Each configuration is also represented by an image taken at an earlier time and a (treated) image taken at later times.}
\label{Fig3}
\end{figure*}

\begin{figure}[h]
\centering
\includegraphics[width=3.45in]{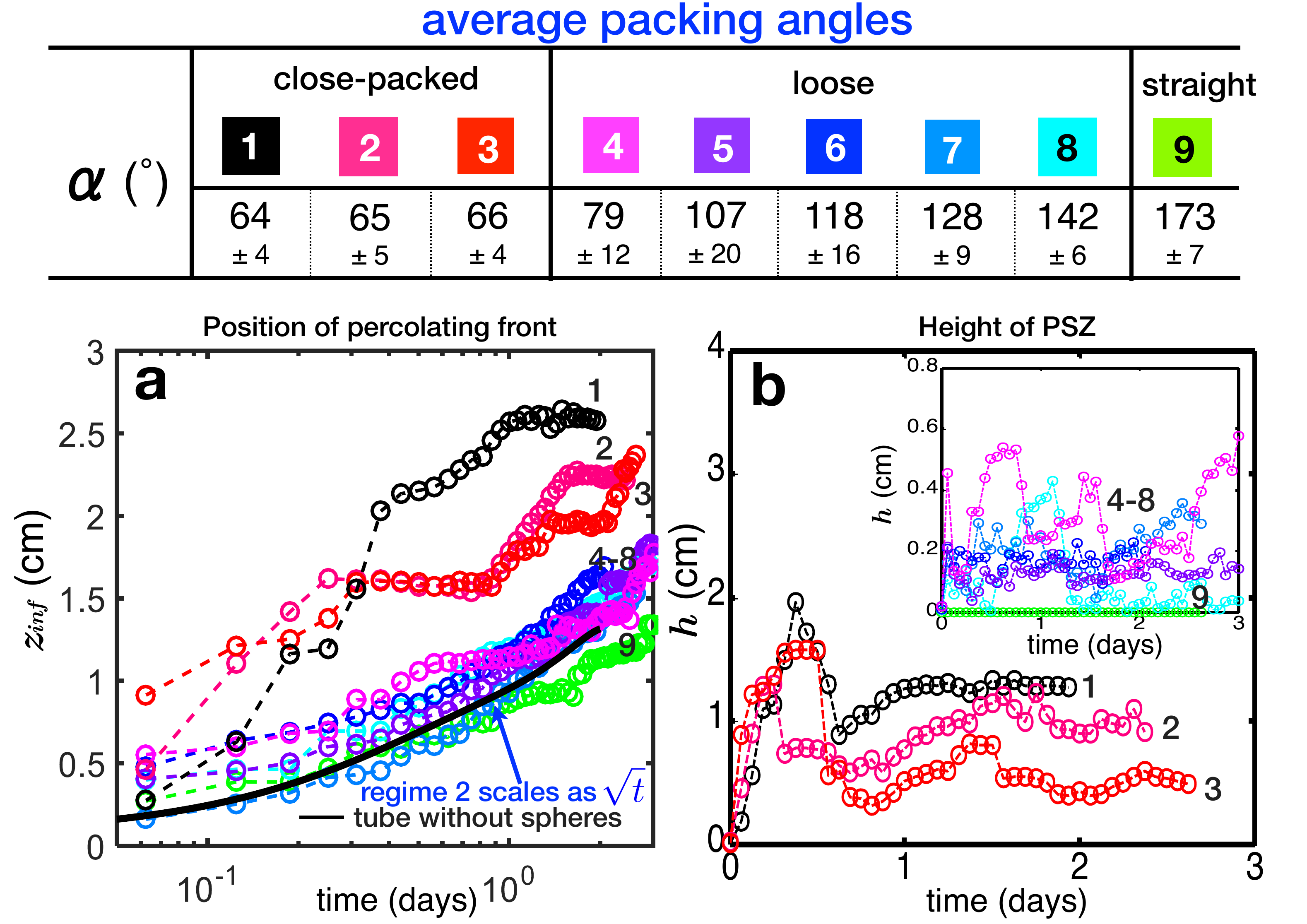}
\caption{(\textbf{a}) Plot of $z_{inf}=\widetilde{z}(t, \alpha)$. Each color corresponds to a packing angle. For close configurations (1-3), we observe a linear regime 1 followed by an abrupt transition to regime 2. For loose configurations (4-8), the curves scale as $z_{inf} \propto \sqrt{t}$, similar to the solid line, which represents experimental results of evaporation from a circular capillary tube (without spheres).  It represents the propagation of the front meniscus as a function of time. (\textbf{b}) Plot of $h=\widetilde{h}(t, \alpha)$. For close-packed (1-3), a maximum is reached before stabilizing at a constant value. For loose-packed (4-8 in inset), the thickness fluctuates suggesting its instability. And for straight configurations (9 in inset), it is nearly inexistent. }
\label{Fig4}
\end{figure}

Capillary tubes with columns of packed beads are initially filled with deionized water up to the brim before starting controlled evaporation experiments. The main phenomena in drying of wet granular media are shown in Fig. 2, with two distinct regimes associated with evaporation~\cite{Coussot00, Lehmann08, Shokri08, Yiotis12b, Yiotis12a, Prat02, Prat07, Chauvet09, Chauvet10, Cejas17} - from capillary-supported nearly constant rates (regime 1) to diffusion-controlled decreasing evaporation rates (regime 2)~\cite{Coussot00, Chauvet09, Prat07, Yiotis12a, Yiotis12b, Lehmann08, Cejas17}. During regime 1, liquid-gas interfaces form that exhibit high capillary pressures, permitting liquid flow from the bulk to the evaporating surface. The films form the Partially Saturated Zone (PSZ) or vadose zone bound by a percolating front~\cite{Yiotis12b, Chaouche94} denoted as  $z_{inf}$ in the ``inferior'' or lower portion and an evaporating front, $z_{sup}$, in the ``superior'' or upper portion. Typical evaporation patterns are shown in Fig. 3, with PSZ images for different arrangements.

We observe different behaviors depending on packing configurations. For straight configuration, Fig. 3a. ($\alpha = 180^{\circ}$), the packing angle only allows a liquid film connection in the order of one sphere and thus, mostly only isolated bridges form. This behavior is observed until  $\alpha \sim 100^{\circ}$, where films subsist in the PSZ in Fig. 3b within a few bead diameters. When the packing angle reaches close-packed configuration ($\alpha \sim 60^{\circ}$), a large PSZ appears in Fig. 3c, where liquid films are attached to the top surface during regime 1. In such a case, hydraulic transport is fast - an action described as a wicking effect~\cite{Cejas17, Yiotis12a, Xu08} - to sustain high evaporation rates~\cite{Coussot00, Cejas17} while $z_{inf}$ continues to recede inside the medium. Recent results~\cite{Cejas17} have shown that the PSZ is in fact a mixture of continuous liquid films and disconnected droplets. Here, we show that straight and loose-packed configurations favor disconnected droplets more than connected films while close-packed configurations favor stable film formation that accelerates drying~\cite{Yiotis03}. The latter thus increases the lifetime of regime 1 while delaying the transition to regime 2.

The qualitative observations are confirmed from the measurements of $z_{inf} =\widetilde{z}(t)$ in Fig. 4a, where the curves are numbered according to $\alpha$ encoded on the legend bar. For close-packed configurations (1-3), the flux is constant in regime 1 and thus the front position, which is the integral over time, scales linearly with time~\cite{Lehmann08, Shokri08, Yiotis12b, Prat02, Coussot00}. The evolution of $z_{inf}$ at shorter times is noiser in close-packed systems due to packing defects, e.g. in curves 2$\&$3 the unexpected presence of a larger pore in the upper parts of the tube allows the front to recede much further. Eventually, it encounters a larger bead that occupies much of the space, thereby temporarily blocking the movement of the receding front and thus breaking the hydraulic connections prematurely. Moreover, when $\alpha$ is increased from close-packed to loose-packed, we observe a drastic change in behavior. For loose configurations (4-9), $z_{inf}$ increases at a very slow rate and thus, there is no observable first regime. In such cases, the films depin easily from the surface and capillary connections gradually thin out~\cite{Coussot00}, thereby decreasing film cross-section and increasing flow resistance from the bulk to the surface~\cite{Chauvet10b, Yiotis12a}. When the films finally detach, a dry region develops - a phenomenon easily observed in straight and loose configurations (Fig. 3a and 3b). This marks the onset of regime 2 of evaporation, where water vapor diffuses across the dry region and thus evaporation rates slow down and $z_{inf}$ scales as $t^{1/2}$~\cite{Camassel05, Bird02, Chauvet09, Yiotis12b, Cejas17}. This behavior is similar to water evaporation out of a circular tube without spheres (solid line in Fig.4a).

This same observation is inferred by looking at the extents of the height of the PSZ, $h=\widetilde{h}(t)$ in Fig. 4b. Despite fluctuations, close-packed systems (1-3) clearly show two distinct regimes, where the height of PSZ increases. The PSZ reaches a characteristic height, determined by the extent of the liquid film region during regime 1. This extent depends on contact angle and pore shape and size. Drying conditions such as lower relative humidities permit transition to occur faster and also decrease the overall water saturation in the medium but does not significantly modify the extent of the PSZ~\cite{Yiotis12a}. The abrupt decrease in height afterwards corresponds to the film depinning from the top surface before $h$ stabilizes during regime 2. For loose and straight configurations (4-9 in inset), the PSZ seems unstable and $h$ heavily fluctuates around a few bead diameters (0.1-0.3~cm) without a distinct linear regime and thus  $z_{sup}$ disconnects immediately from the evaporating surface and the PSZ iz mainly composed of disconnected films. This confirms an unstable stable regime 1 for loose and straight configurations, because the system is instantly dominated by the slow regime 2 of diffusion. 

To summarize, these experiments exhibit similar characteristics as evaporation of a 2D porous medium in Hele-Shaw cells~\cite{Coussot00, Lehmann08, Shokri08, Yiotis12b, Yiotis12a, Prat02, Prat07, Chauvet09, Chauvet10, Cejas17}. With all other parameters being constant, the transition from regime 1 to regime 2~\cite{Chraibi09, Shokri09, Cejas17} in these experiments is dominantly influenced by pore shape~\cite{Prat07, Yiotis12b, Chauvet09, Chauvet10b}, which in this study, is linked to packing angle. Above a certain packing angle, a stable PSZ with connected liquid films cannot form.


\section{Modeling Description}

\begin{figure}[h]
\centering
\includegraphics[width=3.5in]{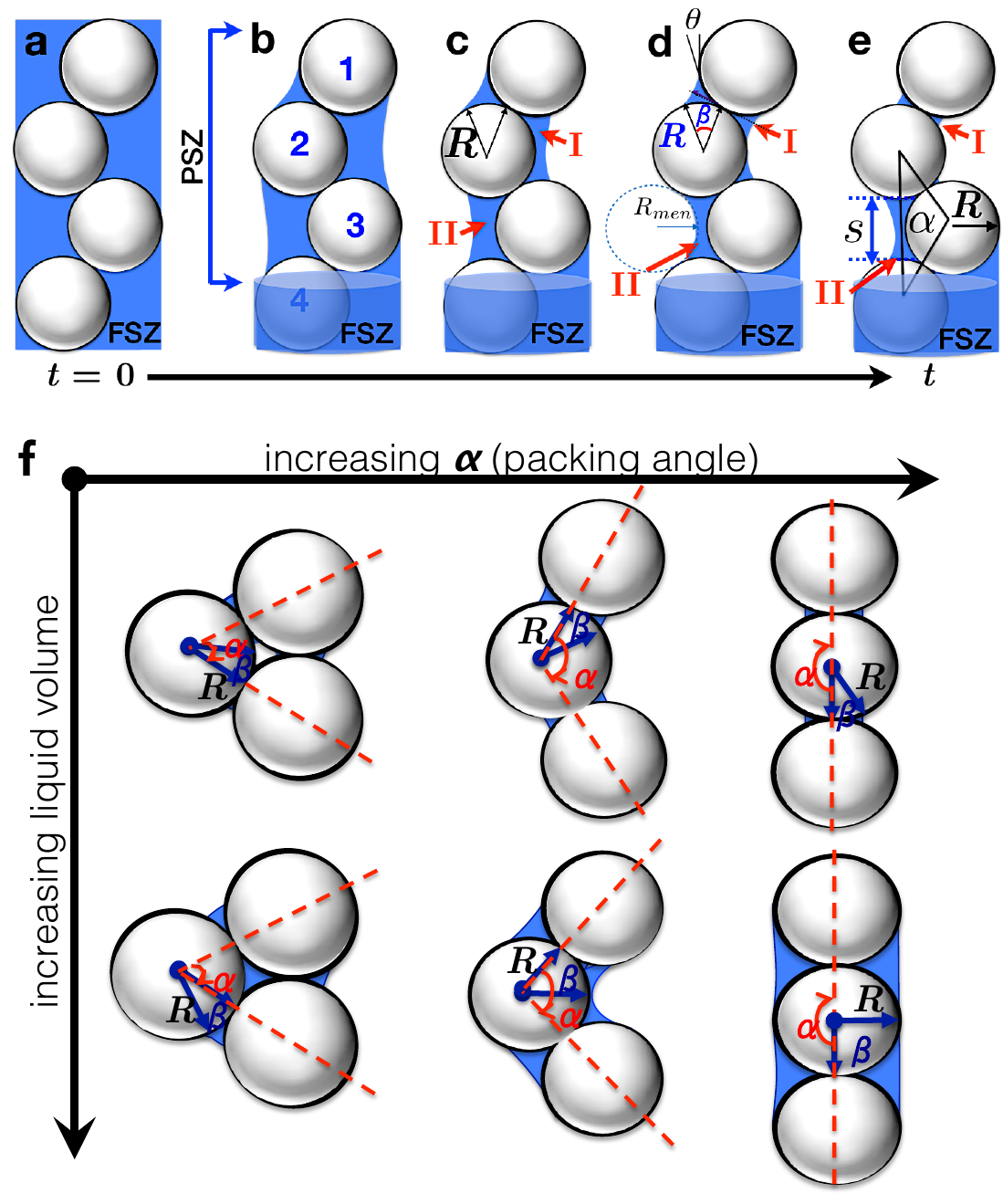}
\caption{(\textbf{a-e}) Schematic of the evolution of an evaporating meniscus for loose configurations. The system starts out with fully saturated zone (FSZ). Initially, liquid films denoted as ``I" and ``II" are connected but as evaporation proceeds, these liquid films break and thus become isolated liquid bridges. The packing angle, $\alpha$, influences the rupture of the menisci connections. (\textbf{f}) Schematic of the effect of liquid volume and packing angle, $\alpha$, on film continuity between three spheres. At constant volume, the liquid body is ``stretched'' and readily separates into two isolated liquid bridges as $\alpha$ increases. At fixed $\alpha$, increasing liquid volume will recover liquid film continuity as depicted by the evolution of the bridge angle $\beta$. The minimum to maintain liquid continuity is $\beta = \alpha / 2$.}
 \label{Fig5}
\end{figure}

\textbf{General description of the model.} We describe the physical process in Fig. 5. The initial configuration (Fig. 5a) shows a fully wet tube with beads. As water evaporates, the volume per unit length of the meniscus decreases (Fig. 5b), air continues to penetrate inside the medium, and the menisci form between spherical beads 1, 2, and 3 (Fig. 5c), with``I" and ``II" being the menisci between beads 1 and 2 and between beads 2 and 3 respectively. At the start, menisci ``I" and ``II" are linked, water from one meniscus can flow into the other. Over time (Fig. 5c), more menisci develop along the tortuous packing and a column of water forms the PSZ, initially composed of connected films. At the top of the column, however, meniscus volume ``I'' decreases until it reaches a point where the connections between ``I'' and ``II'' break (Fig. 5d). After prolonged periods, this connection can no longer exist (Fig. 5e) as the meniscus volume decreases due to evaporation and at this point, the PSZ consists mainly of disconnected bridges.

In close-packed systems, a pore throat exists that is not readily present in loose or straight configurations. In such systems, evaporation proceeds first by loss of liquid volume in the pore throat, resulting to the formation of three liquid bridges of high curvatures, having a larger magnitude of the negative Laplace pressure and thus are able to support a longer film region or PSZ~\cite{Scheel08}. Initially these menisci ensure hydraulic connectivity but as dewetting occurs via evaporation, they eventually disconnect. This observation agrees with literature, where the decay of liquid bridges in close-packed systems is marked by the loss of liquid in the pore throat before the separation into isolated liquid bridges~\cite{Melnikov15, Sembrepon16}.

We show in Fig. 5f a simple schematic on the effect of volume and packing angle. At a fixed liquid volume, larger $\alpha$ (while keeping the other two points of contact) separates the meniscus into two connected bridges. Furthermore, at fixed $\alpha$, the value of the bridge angle or half-filling angle, $\beta$, increases with volume until liquid film continuity is recovered. Thus, the minimum to maintain liquid continuity is $\beta = \alpha / 2$. This is the condition for coalescence~\cite{Scheel08} of capillary bridges for small $B_o$ numbers when the gradient of hydrostatic pressure is insufficient to modify the global shape of the meniscus interface. Greater quantities of liquid volumes increase the value of $\beta$. For close-packed systems of $\alpha \sim 60^{\circ}$, the minimum value of $\beta$ equates to half of the angle of an equilateral triangle~\cite{Scheel08}. Tomography data have indeed shown that liquid bridges in a dense, close-packed system achieve coalescence if $\beta = 30^{\circ}$~\cite{Scheel08, Scheel09, Melnikov15}. 

The condition for de-coalescence, $\beta^{break}$, on the other hand, can occur when $\beta^{(break)} < \beta$, when volume is insufficient, provided that the opening angle provides a good approximation to quantify the shape of the two fused bridges~\cite{Sembrepon16, Melnikov15}. When de-coalsecence occurs even just on one liquid bridge pair, then the liquid film continuity in the PSZ is disrupted. This is true for liquid bodies at $z_{sup}$, where they are more exposed to the atmosphere.

\textbf{Height of liquid film region (PSZ).}  Liquid flow upward towards the top of the tube due to evaporation is driven by a competition between capillary, viscous, and gravity forces. The Bond number, $B_o = \frac{\rho g R^{2}}{\sigma \text{cos}(\theta)}$, where $R$ is the bead radius, $\sigma$ is the surface tension of water, and $\theta$ is the contact angle, gives $B_o \sim 10^{-3}$.  When capillary forces are greater than gravity, the PSZ is formed. In the opposite case, there is no PSZ but rather a flat front between fully wet and dry zones~\cite{Prat02, Prat07}.

The dimensionless film-based capillary number is defined as, $C_a = \frac{3\pi\eta D_M C_e}{\sigma \text{cos}(\theta) \rho_l R}$~\cite{Yiotis12a, Yiotis12b}, where $\eta$ is the dynamic viscosity, $D_M$ is the apparent diffusion coefficient of the vapor through gas phase, $C_e$ is the saturated vapor concentration determined from the saturated vapor pressure at temperature, $T=32^{\circ}$C. We obtain $C_a \sim 10^{-7}$.  Thus, capillary forces dominate over viscous forces during regime 1 and viscous effects only begin to have an effect during the transition from regime 1 to regime 2 at the point of rupture, when the film connections begin to thin out. Overall, the contribution of viscosity is relatively minimal and under these conditions, most of it is also accounted for by balance between capillarity and gravity~\cite{Cejas17}. In addition, surface tension gradients are also minimal because of constant temperature inside the controlled chamber. Note that the walls of the capillary tube are made hydrophobic to limit the wetting effects between the liquid and the tube wall. 

The water content near $z_{sup}$ is lesser than that of $z_{inf}$ at the interface of the fully saturated zone (FSZ). Hence, for a certain height, $h$, the difference in water volume generates a pressure gradient  along the interfaces of the menisci in $z_{inf}$ and $z_{sup}$ of the PSZ that is in equilibrium with the hydrostatic pressure~\cite{Princen69, deGennes10}:
\begin{equation}
 h = \frac{\bigtriangleup P_{inf} - \bigtriangleup P_{sup}}{\rho g},
\label{Eq1}
\end{equation}
where $\rho$ is water density, $g$ is the acceleration due to gravity, $\bigtriangleup P_{inf}$ is the change in pressure at $z_{inf}$ while $\bigtriangleup P_{sup}$ is the difference in pressure between the film and the surrounding air at atmospheric pressure at $z_{sup}$. These two pressures at distinct points in the network are quantities that contribute to the extent of the PSZ. We estimate these pressures through the following analyses.

\textbf{Description of pressure at upper portion of PSZ}. First, $\bigtriangleup P_{sup}$ is the pressure difference between liquid and gas phases at $z_{sup}$, where menisci volumes are smaller. We assume  that the liquid-gas interface is of spherical shape and constant curvature~\cite{Melnikov15, Haines27, Hotta74}, and that there is equal Laplace pressure within the liquid body~\cite{Melnikov15}. If the packing angle is larger, then the trimer (three liquid bridges between three grains) readily dissociates into a dimer or two liquid bridges~\cite{Sembrepon16, Melnikov15}. Thus, the determining shape of the meniscus in $z_{sup}$ is the capillary bridge between two spheres: 
\begin{equation}
\bigtriangleup P_{sup} = P_g - \frac{\sigma \text{cos}(\theta)}{R_{men, cb}}
\label{Eq2}
\end{equation}
where $P_g$ is the gas pressure, $\sigma$ is the air-water interfacial tension, $\theta$ is the contact angle, $R_{men, cb}$ is the radius of curvature of the capillary bridge meniscus at $z_{sup}$~\cite{Melnikov15}, which is associated with both liquid volume and $\beta$. Scheel, et. al.~\cite{Scheel08} analytically reports the Laplace pressure in a capillary bridge between two spheres in contact:
\begin{equation}
\bigtriangleup P_{sup}=P_{cb}=-\frac{\sigma}{R}\left(\frac{\text{cos}(\beta+\theta)}{1-\text{cos}\beta} - \frac{1}{\text{sin}\beta}\right)
\label{Eq3}
\end{equation}

\textbf{Description of pressure at bottom of the PSZ}. The parameter, $\bigtriangleup P_{inf}$, is the pressure difference between liquid and gas at $z_{inf}$, where menisci volumes are higher. While in proximity to the FSZ, the meniscus volume is also large enough to connect a minimum of three beads and form a stable trimer~\cite{Melnikov15} (Fig. 5). Similarly, the shape of the meniscus in $z_{inf}$ also depends on liquid volume (Fig. 5f). The volume (and thus pressure) of the trimer depends strongly on separation distance $s$~\cite{Melnikov15, Willett00} (Fig. 5e), associated with $\alpha$. Thus, the pressure in the meniscus body joining three spheres can be described by:
\begin{equation}
\bigtriangleup P_{inf} = P_g - \frac{\sigma \text{cos}(\theta)}{R_{men, tri}}
\label{Eq4}
\end{equation}
where $R_{men, tri}$ is the radius of curvature of the trimer within three beads, associated with both liquid volume and $\beta$. Here, we also assume that the liquid-gas interface at $z_{inf}$ is of spherical shape and constant curvature~\cite{Melnikov15, Haines27, Hotta74}, and that there is equal Laplace pressure within the liquid body~\cite{Melnikov15}.

Having qualitatively described the forms of the meniscus in $z_{sup}$ and $z_{inf}$, we now quantitatively define $\bigtriangleup P_{sup}$ and $\bigtriangleup P_{inf}$.


\section{Modeling Results and Discussion}
All numerical simulations are performed using Surface Evolver~\cite{Brakke96}, an open source interactive program that studies the evolution of the shapes of liquid surfaces subject to various energies (e.g. surface tension) and constraints (e.g. fixed volumes in units of sphere radii). 

In Surface Evolver, we numerically determine the energy minimum~\cite{Brakke10, Carter96, Wang17} of a meniscus volume between spheres in contact. While it is difficult to determine if in fact this value is the most stable form, using the energy minimum value provides a reasonable approximate. In such calculations, the assumption of hydrostatic pressure is a first-order estimate that does not account for pressure drop due to viscous dissipation, which can result to a modification in the interfacial curvature~\cite{Cejas17, Keita16, Beyhaghi14}. However, preceding discussion shows that the effect of viscous dissipation is minimal when compared to capillary suction.

\textbf{Pressure results at upper portion of PSZ}.  At $z_{sup}$, menisci pressure is described by $\bigtriangleup P_{sup}$ in Eq. 2. De-coalescence readily occurs and the trimer decays into a dimer. In Eq. 2, the calculation for the radius of curvature can be associated to the change of $\beta$ with respect to liquid volume. Results from simulations are shown in Fig. 6.

\begin{figure}
\centering
\includegraphics[width=3.5in]{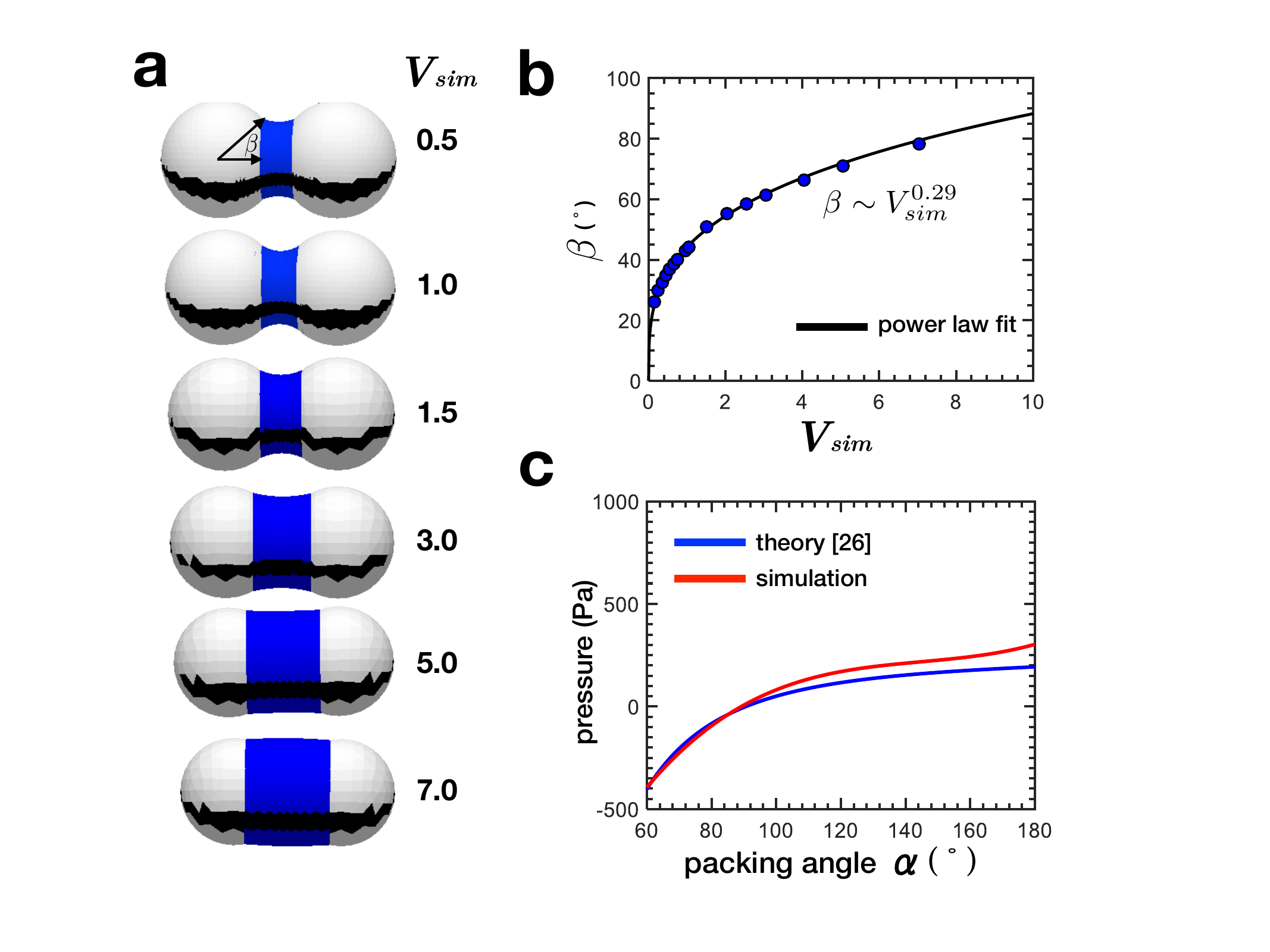}
\caption{(\textbf{a}) Typical examples of Surface Evolver minimization of capillary liquid bridge between two spheres at different menisci volumes. (\textbf{b}) Plot of $\beta$ as function of simulation meniscus volume, $V_{sim}$. The best fit is a power-law of the form, $\beta \sim V_{sim}^{0.29}$. (\textbf{c}) Plot of capillary bridge pressure as function $\alpha$, using the simulation results (Eq. 2) compared with the analytical theory in Ref.~\cite{Scheel08} (Eq. 3). } 
 \label{Fig6}
\end{figure}

We use a self-written code that provides $P=\widetilde{P}(V_{sim})$ at a fixed meniscus volume in the simulation, $V_{sim}$, given surface tension ($\sigma=0.072$ N/m) and prescribed contact angle. We obtain the pressure value from the derivative of energy with respect to volume,  $P_{sim} = dE/dV_{sim}$. The spheres used in the simulation are in contact with one another. In this case, we vary $V_{sim}$ in calculating the pressure of the meniscus body joining two spheres. The condition of ``two spheres'' is important at $z_{sup}$, for reasons mentioned in the previous section. Pressure values calculated from Surface Evolver~\cite{Brakke10, Carter96}, $P_{sim}$, are normalized using: $P= (1/\lambda) P_{sim}$, where $\lambda$ is a rescaling factor taking into account the length scale used in the simulation and the actual bead size used in the experiment. Similarly, $V_{sim}$ can be normalized using $V = V_{sim}\lambda^3$. We then determine the relationship between $V_{sim}$ and $\beta$, simply by taking the vertex at the point where the liquid meniscus touches the sphere, of which some simulation examples are shown in Fig. 6a. We find: $\beta \sim V_{sim}^{0.29}$ for the meniscus between two spheres (solid line in Fig. 6b). Rabinovich, et. al.~\cite{Rabinovich05} shows that for a liquid bridge between two spheres in contact, $\beta \sim V_{sim}^{0.25}$. The slight discrepancy of our exponent is due to the sensitivity of the analysis to contact angle values, which in our simulation is different with respect to the one reported in literature~\cite{Rabinovich05}.

Nevertheless, to further validate our Surface Evolver approach, we compare the results of our simulation with that of the theoretical analysis proposed by Scheel, et. al.~\cite{Scheel08} in Eq. 3. From Surface Evolver, we determine the pressure of a meniscus volume and using the $\beta \sim \widetilde{\beta} (V_{sim})$ relationship (Fig. 6b), in addition to $\beta=\alpha/2$, we take a least squares polynomial fit of the simulation results $\bigtriangleup P_{sup}=\widetilde{P}(\alpha)$ and plot it against Eq. 3. The result is shown in Fig. 6c, demonstrating decent agreement between the simulations and the analytical equation~\cite{Scheel08}, especially at lower values of $\alpha$. The slight deviation at higher values of $\alpha$ is insignificant considering that experimental results (Fig. 4) show minimal if not inexistent $h$ values at $\alpha>80^{\circ}$.

The agreement between Surface Evolver and reported analyses from literature justifies our numerical approach.

\textbf{Pressure results at bottom portion of PSZ}.  To calculate $\bigtriangleup P_{inf}$ in Eq. 4, we also use a slightly modified Surface Evolver code originally written by Carter~\cite{Carter96} that directly provides $P=\widetilde{P}(\alpha)$ at a fixed meniscus volume, $V_{sim}$ given surface tension ($\sigma=0.072$ N/m) and prescribed contact angle. Similarly, we obtain the pressure from  $P_{sim} = dE/dV_{sim}$. The middle sphere used in the simulation is in contact with the two other spheres. In this case, we vary $V_{sim}$ and $\alpha$, in calculating the pressure of the meniscus body joining three spheres. The condition of ``three spheres'' is important in $z_{inf}$ for reasons mentioned in the previous section. In this simulation, the middle sphere remains fixed while two spheres at either ends are gradually separated by $\alpha$. Similarly, $P_{sim}$ values are normalized using: $P= (1/\lambda) P_{sim}$. Likewise, $V_{sim}$, can be normalized using $V = V_{sim}\lambda^3$.  For a given $V_{sim}$, we vary $\alpha$ shown in Fig. 7a.  For a certain constant $V_{sim}$, the pressure becomes increasingly negative as the meniscus is ``stretched'' when the angles open up. 

For a given value of $\alpha$ in Fig. 7a, we take the corresponding values of pressure for different $V_{sim}$ and replot them as a pressure-volume curve in Fig. 7b. The general trend in Fig. 7b is that pressure becomes increasingly positive at higher volumes - an observation supported by the inverse relationship between pressure and volume, $i.e.$ lower volumes exhibit increasingly (in terms of magnitude) negative pressures~\cite{Wang17}. Figure 7c shows the same pressure-volume curve but with additional interpolated points within the given volume range. Finally, images in Fig. 7d show typical examples of Surface Evolver minimization for meniscus in three spheres at different $\alpha$ but for constant $V_{sim}$. Other simulations are also performed at different values of $V_{sim}$ but at constant $\alpha$. We obtain no numerical solution for $\alpha > 120^{\circ}$.

\begin{figure*}[h!]
\centering
\includegraphics[width=5.3in]{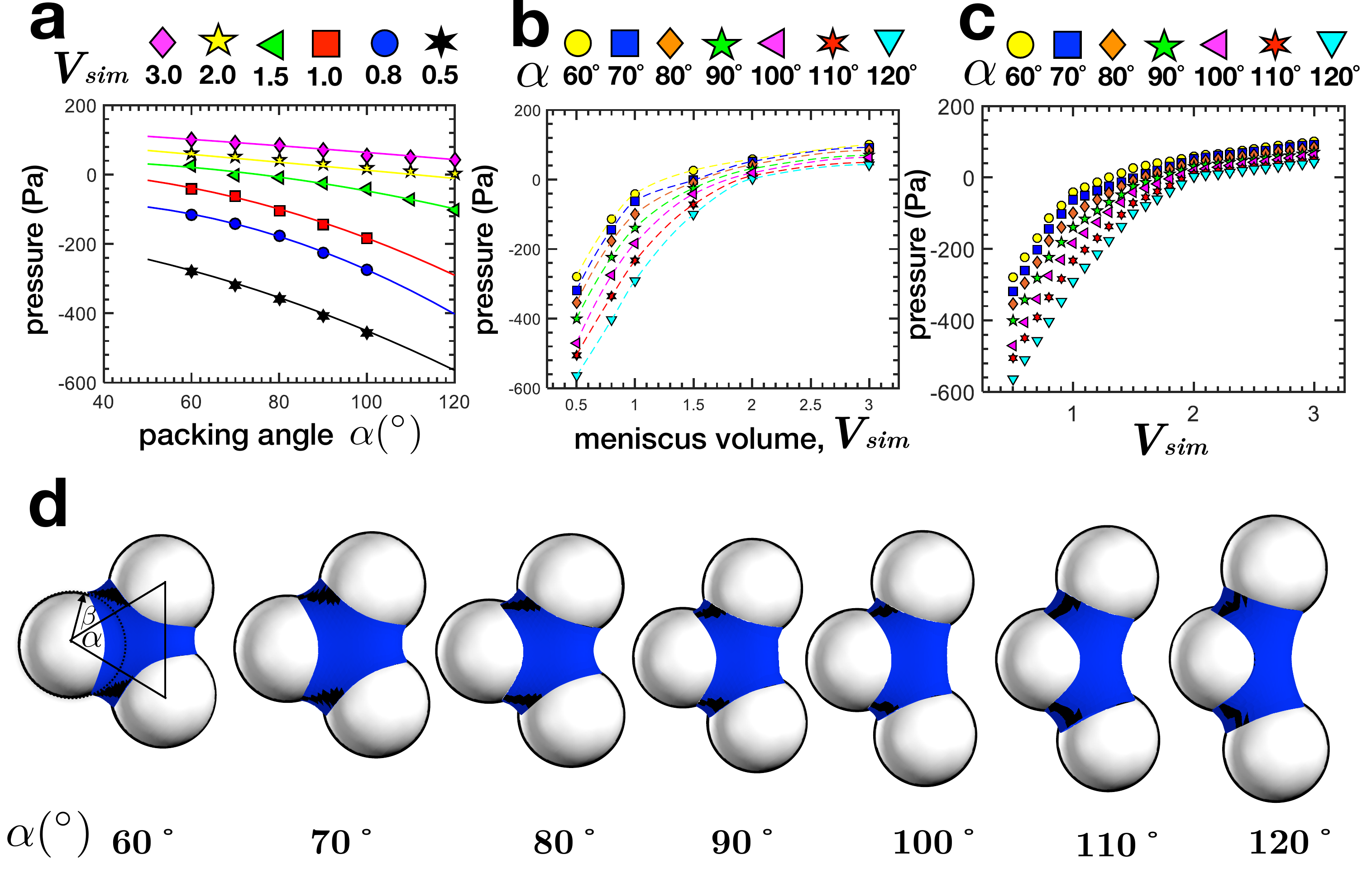}
\caption{(\textbf{a}) Pressure-packing angle relationship for a given $V_{sim}$ as calculated from Surface Evolver. Solid lines are guides for the eye. (\textbf{b}) Pressure-volume curves at a given $\alpha$ (units in degrees). Broken lines are guides for the eye. (\textbf{c}) Pressure-volume curve with interpolated points within the given volume limits. (\textbf{d}) Typical examples of Surface Evolver minimization of a constant meniscus volume, $V_{sim} = 1.0$, at different $\alpha$. Surface Evolver calculates the shape of the meniscus within three spheres (in blue). The spheres in the code~\cite{Carter96} are transparent so only the meniscus is visible during iteration and thus spheres were later superposed for better visualization of the meniscus.} 
 \label{Fig7}
\end{figure*}

\begin{figure*}[h!]
\centering
\includegraphics[width=5.3in]{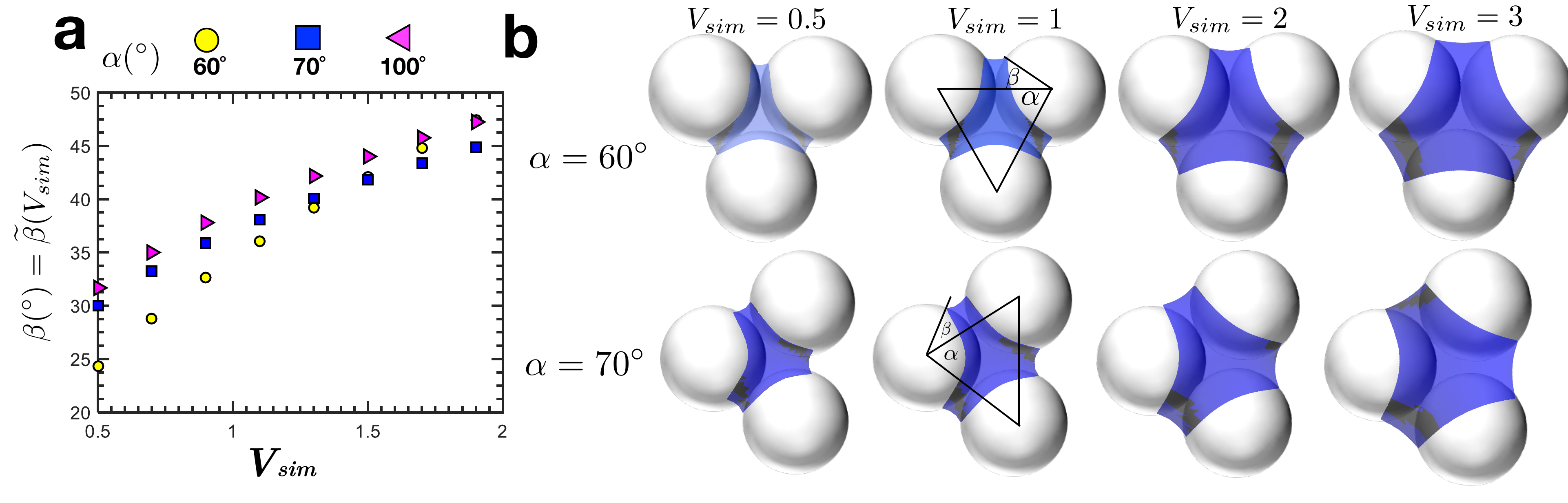}
\caption{(\textbf{a}) Plot of $\beta$ as function of increasing $V_{sim}$ at constant $\alpha$. The best fits are described by a power law of the form, $\beta \sim V_{sim}^{0.5}$ for $\alpha \sim 60^{\circ}$ and $\beta \sim V_{sim}^{0.3-0.4}$ for $\alpha \geq 70^{\circ}$. The critical volume corresponds to $\beta=\alpha/2$. (\textbf{b}) Some examples of Surface Evolver for meniscus in three spheres at constant $\alpha$ but varying $V_{sim}$. Similarly, the spheres in the code~\cite{Carter96} are transparent so only the meniscus is visible during iteration and thus spheres were later superposed after iteration for better visualization. Furthermore, the meniscus color has decreased opacity in these examples to better emphasize $\alpha$.} 
 \label{Fig8}
\end{figure*}

For a given $\alpha$, there exists a critical volume of the meniscus body in which film continuity is still maintained. This critical volume can be determined from $\beta$, which also depends on $\alpha$. The minimum value of $\beta$ such that film continuity can exist is $\beta = \alpha/2$ and the critical volume is the volume that corresponds to that value of $\beta$. Similar to the previous section, we determine the relationship between $\beta$ and $V_{sim}$ for a meniscus in three spheres from simulations. For a given constant value of $\alpha$, we vary $V_{sim}$ and similarly take the values of the vertices at the point where the liquid meniscus touches the sphere to determine $\beta$. We perform this for a constant value of $\alpha$ and determine results $\beta=\widetilde{\beta} (V_{sim})$ shown in Fig. 8a. We also show some Surface Evolver examples at constant $\alpha$ but varying $V_{sim}$ shown in Fig. 8b. For every curve of $\alpha$, we take a power law fit of $\beta=\widetilde{\beta}(V_{sim})$. From simulations, the general form for $\alpha \sim 60^{\circ}$ is $\beta \sim V_{sim}^{0.5}$. But from $\alpha \geq 70^{\circ}$, the general form becomes $\beta \sim V_{sim}^{0.3-0.4}$.
 
Although the capillary force of a liquid bridge between two beads has been a well studied classical problem~\cite{deGennes10, Scheel08, Willett00, Rabinovich05}, the problem becomes analytically more complex when three or more spheres are involved. This is where numerical simulations of energy-volume help alleviate the dilemma, as also demonstrated from literature results~\cite{Wang17, Sembrepon16, Melnikov15}. 

\textbf{PSZ height - packing angle relationship}. After having calculated the pressures at both $z_{sup}$ and $z_{inf}$, we solve for $h$ in Eq. 1 and we find numerically that above a certain $\alpha$:
\begin{equation}
\bigtriangleup P_{sup} > \bigtriangleup P_{inf} 
\label{Eq5}
\end{equation}
which is true at higher $\alpha$ values. This might seem counterintuitive since increasingly negative pressures favor capillary suction~\cite{Xu08}. However, this could also suggest that the menisci volumes have become much smaller, and therefore have been disconnected. Thus, instead of joining spheres that favor film continuity, the menisci have decayed into individual, separate bridges. Solving for $h$ with the criterion in Eq. 5 will result to $h < 0$. A negative value for $h$ is physically impossible and is interpretated as a condition when the liquid film network is not stable anymore and thus cannot exist continuously as one single film network. Therefore above a critical angle, when the model predicts a negative $h$ value, all values of $h$ are instead represented as $h=0$. Furthermore, no simulation results (for meniscus in three spheres) were obtained for $\alpha>120^{\circ}$, but it is however highly unlikely that the film network would still be stable and connected at such extremely high opening angles and thus $h=0$ for $120^{\circ}<\alpha \leq 180^{\circ}$.

This ``critical" angle was also observed in our experiments, where no stable film network is formed but instead disconnected droplets and thus only the second evaporation regime appears. We compare the model with our experimental results, as shown in Fig. 9, with the critical angle appearing at $\alpha_{crit} \sim 70^{\circ}$. Results show agreement between experiments and simulations. Experiments at close-packed systems, however, seem to be underestimated, which can be attributed to packing defaults, which augments the risk of disconnected films.

Recent studies~\cite{Melnikov15, Scheel09, Sembrepon16} have noted two cross-overs in the stability of trimers in close-packed dense systems. The first cross-over is described when trimers decay into three liquid bridges at $60^{\circ} < \alpha < 64^{\circ}$ due to bursts that result to an opening of the pore throat~\cite{Sembrepon16}. A second cross-over occurs when trimers decay rapidly into two liquid bridges at $64^{\circ} < \alpha < 68^{\circ}$ described by snap-in instabilities~\cite{Sembrepon16}. At opening angles beyond $\alpha > 68^{\circ}$, this marks the onset of a discontinuous meniscus jump~\cite{Melnikov15, Scheel09}. This suggests that there is no specific angle for trimer break-up and that break-up will occur if the pressure is below a certain critical value. From our results, the ``critical'' angle we obtain falls within the onset of a discontinuous meniscus jump, which is aided by the opening of $\alpha$ or in other words, the ``loosening'' of the granular chains. The larger the value of $\alpha$, the discontinuous jump occurs rapidly. This work shows that increasing the packing angle is similar to increasing the associated overall pore size of the granular chain. It also provides evidence of the fact that a minimal increase in packing angle above the critical angle corresponds to a very large change of the overall pore size.

\begin{figure}[h]
\centering
\includegraphics[width=3in]{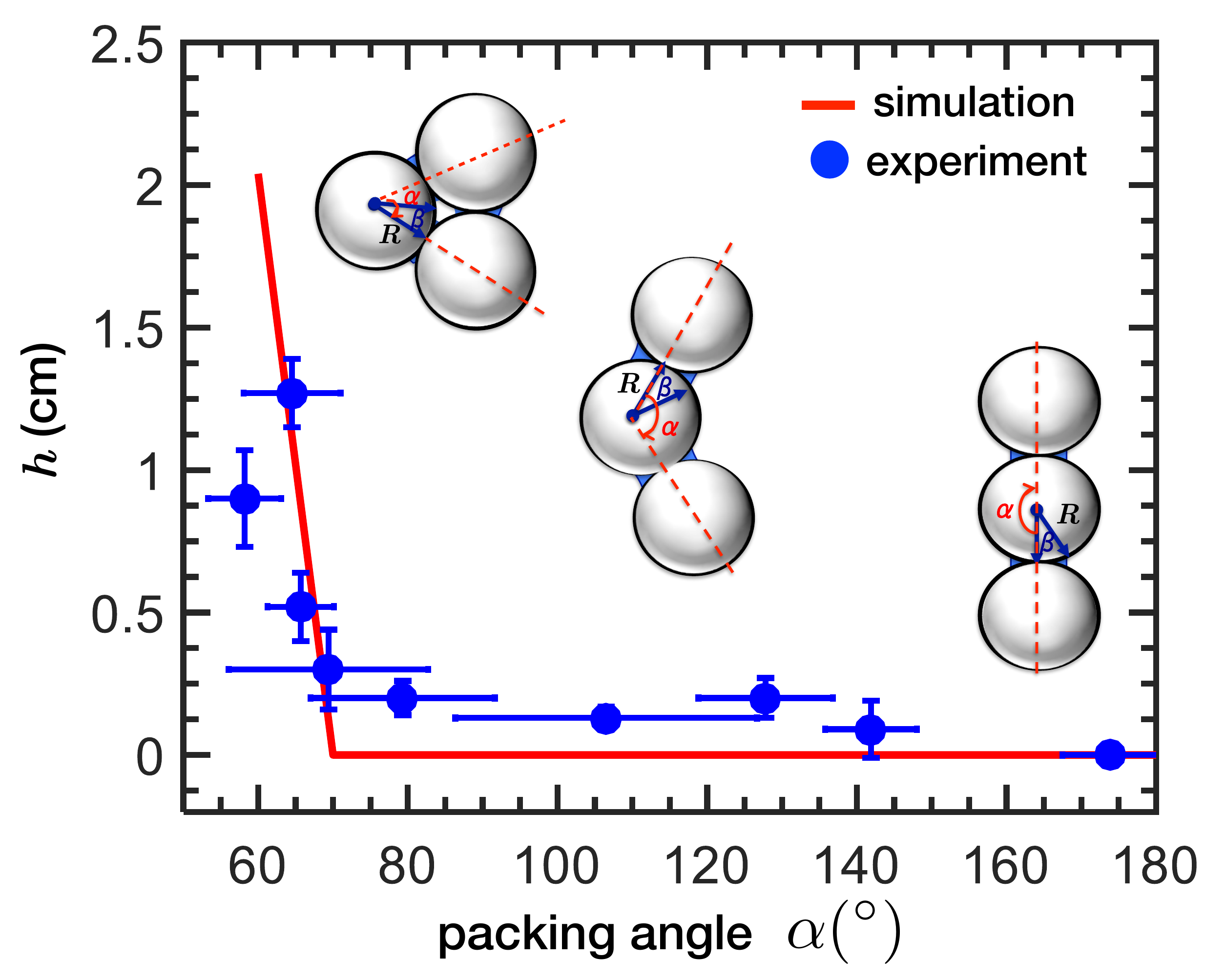}
\caption{The experimental values ($\bigcirc$) obtained for $h=\widetilde{h}(\alpha)$. The solid line is the result of Eq. 1. The sharp transition of the curve shows that stable connections are quickly broken once the granular column is loosened. }
 \label{Fig9}
\end{figure}

\section{Conclusion}

We experimentally investigated how arrangement or the geometry of the granular chains affects liquid film stability in capillary networks.  Using a simple model system, we varied grain configurations from loose-packed to close-packed as determined by a packing angle. We compared experimental results with Surface Evolver simulations. Using simple geometrical argument, we calculated the pressure contributions and extract the height of the PSZ from these values. Close-packed systems have more stable liquid film connections; thus, evaporation rates are faster.  At a certain packing threshold (loose-packed systems), liquid connections are unstable and they readily break thus having slower evaporation rates. 

The arrangement of particles in bulk systems clearly affects compactness and liquid distribution. Less liquid will be found in densely packed areas since a stable PSZ with mainly connected films favors evaporation. The continuous liquid flms in the PSZ drive water from the bulk to the evaporating surface. In contrast, loosely-packed areas will hold greater liquid content for longer periods because liquid connections quickly break and thus evaporation rates are rapidly driven by vapor diffusion, which are slower than rates driven by capillary action. The results presented in this paper corroborate literature on the consequences of having connected liquid films~\cite{Yiotis03} and disconnected pockets and isolated bridges on evaporation~\cite{Cejas17}.

The results also raise interesting questions on the role of polydispersity on compactness in multiphase flow especially when smaller particles fill the pore spaces made by larger ones. Such configurations create smaller pores, whose liquid films have extremely small curvatures with increasingly negative pressures. The effects of these features are not only observed in evaporation but also in other fluid fluxes occuring the granular media such as infiltration, drainage, and channel formation~\cite{Cejas14a, Wei14}. Such investigations help contribute to the better comprehension of the complex nature of multiphase flow in 3D systems.


\section{Acknowledgements}

We thank the support of Centre National de la Recherche Scientifique (CNRS) Complex Assemblies of Soft Matter (COMPASS UMI 3254) laboratory with University of Pennsylvania. We also thank the CIFRE program of the Agence Nationale Recherche Technologie (ANRT). We thank Jennifer Reiser, Marine Truchet, Alban Sauret, Cyprien Gay, Douglas J. Durian (UPenn),  W. Craig Carter (MIT), and Kenneth Brakke (Susquehanna University) for Surface Evolver discussions and/or comments on manuscript revision.


\bibliography{References_geometrypaper} 
\bibliographystyle{rsc} 

\end{document}